# Kinship Is a Network Tracking Social Technology, Not an Evolutionary Phenomenon

Tamas David-Barrett

Email: tamas.david-barrett@trinity.ox.ac.uk

Address: Trinity College, Broad Street, Oxford, OX1 3BH, UK

Web: www.tamasdavidbarrett.com

**Abstract**

On one hand, kinship is a universal human phenomenon that tends to align with biological relatedness, which might suggest evolutionary foundations. On the other hand, kinship has exceptional variation across the human populations, which points to cultural foundations. Furthermore, even if its foundation was biological, kinship is often too imprecise to track genetic relatedness efficiently, while inclusive fitness theory would suggest focusing only on the closest relatives, which is not the case in most human cultures. It was the parallel validity of these contradicting arguments that led to decades of fierce debate about the definition and measurement of the phenomenon. This paper offers a new approach to kinship. First, the model demonstrates that it is possible to generate kinship networks (a) derived from the kind of basic kin connections that our species shares with other apes, but (b) driven by network rather than biological logic beyond the immediate family. Second the model demonstrates that kinship as a network heuristic works efficiently only in high fertility societies, and gives way to similarity-based friendship with demographic transition. The results explain (i) why kinship labelling is unique to our species, (ii) why kinship is universal among human cultures, (iii) why kinship terminology systems are varied across cultures, (iv) why linguistic kin assignment is imprecise, and (v) why kinship is replaced by homophily when relatives are scarce. The model offers a unifying framework to the debate between social and evolutionary anthropology concerning the concept of human kinship.

# Introduction

Kinship, i.e., a linguistic assignment of relatedness degree among members of a social group, is both unique to our species, and universal among human cultures [1-3].

Although kinship is perhaps the most studied anthropological phenomenon [4-7], its role within behavioural science has been controversial for five decades [8-11]. On one hand, every human culture has some form of kinship concept, reflected by the use of kin terminology that assigns a degree of relatedness using language [12]. This universality would point towards a shared evolutionary foundation. On the other hand, there is an extreme variation among human cultures in terms of the kinship terminologies and practices [13]. This suggests that cultural dynamics would drive the differences between kinship systems. Thus, the universal use of kinship systems suggests that kinship is an evolutionary phenomenon, while the exceptional variation suggests that it is a purely cultural one.

These two seemingly contradictory observations might be brought together by the assumption that while any one particular manifestation of kinship systems is specific to a given human culture, there is an underlying process that is shared among all humans, and it is this dynamic that drives the phenomenon. An obvious candidate for a trait that would serve as a foundation for all human kinship systems is genetic relatedness, which is the general assumption of evolutionary anthropology [14, 15] and evolutionary psychology [16, 17]. However, there are two problems with the suggestion that genetic relatedness is the logic of kinship labelling.

First, as it has been demonstrated by a multitude of studies in social anthropology, kin labels almost never track biological relatedness perfectly [8, 18, 19]. Even when the kinship system corresponds to the genealogical lineage, it tend to make "small errors" in terms of biological relatedness. For instance, the term "cousin" in English covers a range of people with different relatedness coefficients. Yet, while the concept of extended kinship can point to a wider group in which the exact degree of relatedness might not be tracked, the underlying relationship assumed is still the assumption of direct genetic links [4, 20] or indirect ones via shared genetic interests [3, 21], or a construct that uses existing kinship language but moves away from it [22-24]. However, in some cases, kinship is entirely imaginary [25, 26]. This suggests that even if there is a link between kin terminology and genetic relatedness, it is unlikely to be a simple translation from one to the other.

Second, although genetic relatedness has been demonstrated both theoretically [27, 28] and empirically [29, 30] to be the central principle of most forms of costly, altruistic, cooperation in biology, the logic of inclusive fitness suggests that resources should be invested in the closest relatives when there is a choice. Thus, while the theory of inclusive fitness explains the presence of altruistic behaviour as a form of cooperation based on genetic relatedness, unless the fitness returns to increased resources are steeply diminishing, there is no evolutionary reason to help distant relatives when near ones are available [31]. In other words, as long as there are siblings around, the inclusive fitness-based argument would likely leave the cousins, and everyone beyond, ignored.

This raises the question of why there might be a language-based kin tracking system the first place. There is no need for linguistic kin assignment for the recognition of the closest relatives of the same generation, i.e., siblings, as it is covered by the genetically inherited form of kin recognition by developmental proximity [32-34], which humans share with other apes [35-38]. This effect, first discovered by Edvard Westermarck, ensures that siblings who grow up together cooperate as adults without the need to use special linguistic terms.

Thus, the observations that kinship mapping is always imprecise, and that, in any case inclusive fitness would suggest a focus on only the closest relatives, imply that genetic relatedness is unlikely to be the ultimate reason why kinship systems exist.

This leaves us with the question: why does this ubiquitous behaviour, at the same time both unique to humans, and universal among humans, exist?

In this paper I offer a hypothesis to answer this question: kinship is a social network tracking heuristic that is linked to genetic relatedness only via its base being the close-kin, highly related dyads. In other words, if a social network is constructed on the basis of either sibling or parent-child social network edges, then the degree of relatedness is a proximate measure of the distance in the social network, and, as I will show, it is also a measure of the likely cooperative stance towards each other. What this model shows is that kinship network is indeed about cooperation, but along the logic of a social network and not along the logic of shared genes. Thus, it is the *presence* of the network that is based on relatedness, while cooperative stance is based on the structural properties of the network.

The insight is the following: if person A is more closely related to person B than to person C, then, by definition, the number of shared relatives between A and B is higher than the number of shared relatives between A and C. Thus, if a social network consists of only sibling or first-degree cousin edges, then A will have a higher number of closed social network triangles with B than with C.

Let us assume that A needs to choose either B or C too have a cooperative act with, and that A does not have a past experience with either potential partner, or any other information, apart from the degree of relatedness with B and C. Let us also assume that the cooperative act is such that each side can cheat or not on the other, e.g., in a prisoner's dilemma setup. If one cheats, it is costly for the other side.

If person A does not want to be cheated on, she needs to choose a partner who she can trust. If there is a space to gossip, then she knows that the only reason for B or C not to cheat on her is their worry that if they do so, she will tell their shared connections about this fact. This network reputation effect is stronger the more shared connections they have [39]. She knows this, and thus for her it makes sense to choose to cooperate with the person with whom she has a higher number of shared connections.

**Hypothesis 1.** If all social network connections in the social network are kin connections, then the degree of relatedness is a good predictor of the number of shared connections, and thus an effective way to inform A that she should cooperate with B rather than with C. In this sense, the degree of relatedness can be seen as merely a way to keep track of the likely number of shared social connections with others, and thus a predictor of their cooperative stance.

This hypothesis originates from the existence of a link between kinship tracking and social networks during demographic transition [40-42], which shows that the micro-structure of the social network is dependent on the availability of relatives.

**Hypothesis 2.** This mechanism that links kinship with cooperation via the social networks, suggests a second hypothesis, as a consequence: kinship degree will be a useful predictor of the cooperative stance of potential partners only if there is a sufficient number of relatives available with whom to populate the social network. Hence, kinship matters, i.e., it is a useful mental tool of relative social network position accounting, but only in high fertility, rural, non-migratory societies, and vanishes in low fertility, urban, migratory societies.

**Structure of the paper.** The first section is an illustration for how linguistically assigned kinship networks can be derived from biologically inherited kin-recognition. This is followed by a non-mathematical description of the model and the results. The detailed, formal maths can be found at the end of the paper as Mathematical Methods.

# How to build a kin network?

It is not only our species that uses kin networks to coordinate what the group does. Many animals use relatedness as an organising principle for their societies [29, 43].

Social insects, such as bees, wasps, termites, ants, all use close relatedness cues when deciding if an individual is part of the hive or nest [44]. Although similar in some respect, these are not kin networks in the way we see them in humans. As all those who have seen an ant nest or a beehive closely know well, these might have an internal structure, separating groups into specialised hive "organs", each focusing on a different set of tasks, yet these do not process a network structure among individuals. The reason is simple: for a network structure of any stability to last it is essential that individual edges also last. These social insects divide task not based on social network position of individual group members, but according to genetically defined types and tasks. Cooperation within a beehive is more similar to the human body than to the human society.

It is possible to create a basic kin network based on simply on who is around. Many birds use proximity as a guiding principle with whom to collaborate. If an individual from these species meet another more frequently, then it is more likely to engage in cooperation, than with another bird, further away from the nest and hence less frequently met [45]. The logic is simple: these species tend not to disperse much from the parental nest, and thus their own nest is bound to be close to others from coming from the same parental nest, or the nest of the parent's siblings. The more frequently one bumps into the other bird, the higher is the likely relatedness between them.

It is even possible to build complex, multi-level societies using this 'method' of proximity, as long as the individuals are stationary in their positions compared to other group members, defying the need for high computational power. Coordination of some small-brained bird networks emerges this way, for instance [46]. However, proximity-based kin recognition is not a particularly precise method. The more widely members of a species disperse the less useful this tool is.

Apes, including humans, cannot build kin networks based on genetically defined types and tasks, as there are none, and we move about too much for proximity to be a useful kin marker. Instead, in our branch of evolutionary tree, another solution has evolved. It has three key characteristics: (1) each individual can tell the other individuals apart from the others group members, and (2) they have good memory to remember each other. For these, large brains are needed [47-49]. If childhood is long, then for kin recognition all the individuals need to do is remember those they grew up with. In other words, (3) widely dispersing species with good memory can use developmental proximity as a rule that decides who is related [50].

Using this capacity, building a social network of relatives is straightforward. For example, let us assume that a family of two heterosexual parents, with zero paternity uncertainty, has three children. Each pair of the three children will have a Westermarck-dyad between them, and each child will have one with both parents (Fig.1).

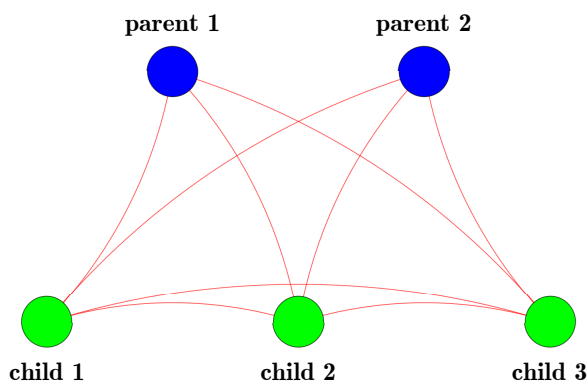

Fig. 1. Development proximity-based, i.e., Westermarck-dyads of kinship within a two-generational family of two parents and three children. Each red line represents a kin relationship that is recognised based on living in the same family during childhood.

Let us assume that every family in the society has the same pattern (Fig. 2). Note that this graph from the point of view of the children's generation.

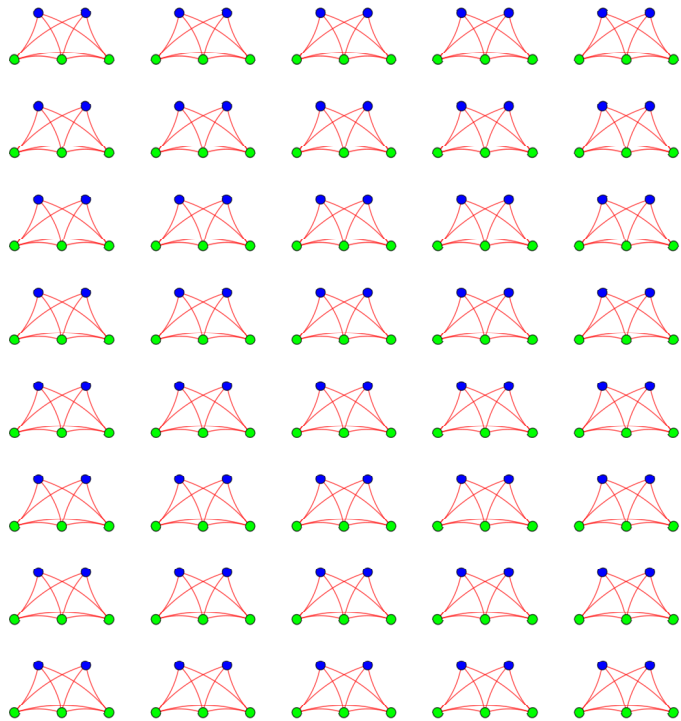

Fig.2. The Westermarck-dyad network of a population of 40 two generation families. Blue: parents, green: children.

The previous generation, i.e., that of the parents', also had their own siblings, and thus their Westermarck-dyads. Assuming that the parents' parents also lived in the two parent, three children family pattern, this additional step allows us to construct a network of the entire population for two generations. This gives us a full network defined by Westermarck-dyads: those between the children, those between the children and their parents, and those between the parents and their respective siblings (Fig. 3). Note that the graph depicted is an illustration: the particular parents' sibling allocation was generated randomly, and is one instance of a high number of possible random instances.

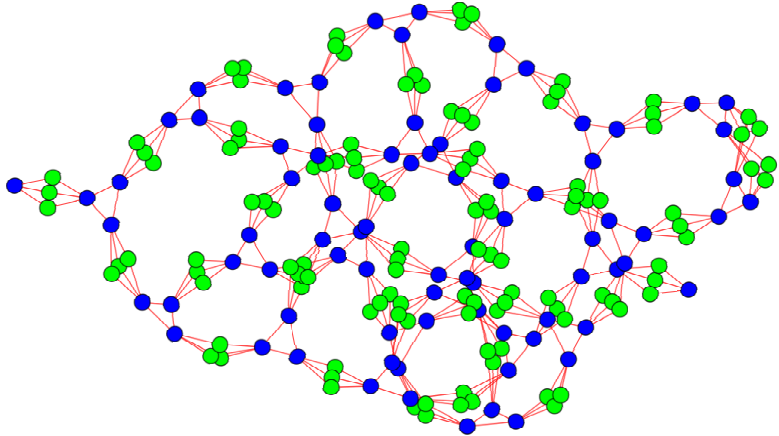

Fig. 3. Example kin network of a society entirely consisting of Westermarck-dyads. Blue: parents, green: children. Red line: recognised kin network edge based on developmental proximity.

The network of Westermarck-dyads can be used to assign kinship using language labels based on the number of edges between two individuals. For instance, a person labelled as "first-degree cousin" can be defined as a three-edge network path: from ego to ego's parent (edge 1), to ego's parent's sibling (edge 2), to ego's parent's sibling's child (edge 3). Thus, all first degree cousins will be exactly three steps away. Similarly, the number of steps to a second-degree cousin is six. The assignment using language then allows us to drop the parents' generation from the graph and focus on a single generation, i.e., that of the children (Fig. 4).

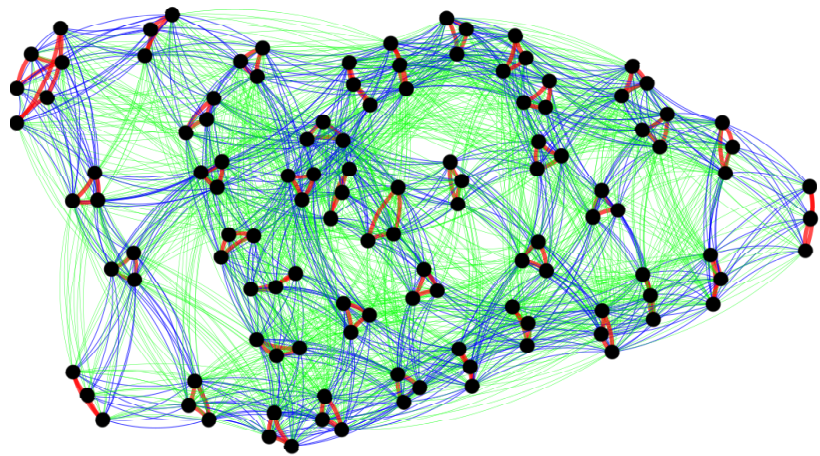

Fig. 4. Example kin network of a single generation. Each black dot represents a person belonging this generation. Red lines: kinship recognised via developmental proximity, blue lines: kinship of first-degree cousins recognised linguistically, and green lines: kinship of second-degree cousins, also recognised linguistically.

Thus, the two-generation graph that consists only Westermarck-dyads (i.e., Fig. 3) can be transformed into a single-generation kin network in which first and second (and any other) degree cousin edges are added to sibling edges (i.e., Fig 4). The linguistic kin assignment defined this way thus still follows the Westermarck-dyads, although indirectly. However, once the assignment is done using linguistic terms, a space for infinite flexibility emerges.

Note that this is the point where social anthropology and evolutionary anthropology tend to clash. The social anthropology tradition focuses on the fact that the most kin relationships in Fig. 4 are assigned by language, and thus can be plastic. The evolutionary anthropology tradition focuses on the fact that in its pure form the linguistic terms are simply translations of genetic relatedness. In the above framing, both traditions are right, and yet they are not necessarily contradicting each other.

This derivation of the kin network shows two key characteristics of human kinship. First, it demonstrates how kinship can be linked to biological relatedness. Second, it shows where the flexibility is: there is no pre-determined obligation for the linguistic kin assignment to follow exactly the relatedness network. Instead, kin labelling is a network tracking method. As the model described in the next section shows, whether kinship is the best technique to track the social network depends on a key assumption: there are available kin around.

# Overview of the model and results

(This section provides a non-mathematical description the model, and presents the model's results. For the step-by-step algebra, see the Mathematical Methods section at the end of the paper.)

In the first step, I built 400 societies on the computer, each with different population histories. Every society grew from a small number of agents to the population size of 2000 individuals. In each generation, women and men were randomly allocated into heterosexual life-long couples, making sure that they are neither siblings nor first-degree cousins. The only variable that differed among the 400 random societies was the fertility of the society, which ranged from 2.5 to 5.0 child per woman.

In these simulations, I tracked the population history, and thus it is possible to calculate the relatedness among the members of any particular society. For ease, I used the number of shared great-great-grandparents as a measure of relatedness. (Every agent has 2 parents, 4 grandparents, 8 great-grandparents, and 16 great-great-grandparents. Siblings share all 16 of their great-great-grandparents, first-degree cousins share 8, second-degree cousins share 4, and third-degree cousins share 2.) When the population size reached the limit of 2000 people, I stopped the simulation, and recorded the relatives for everyone.

Fig. 5 shows the number of relatives an average person in societies of different fertility. Notice that fertility drives histogram: in high fertility societies, the bulk of the generation has shared great-great-grandparents, whereas in low fertility societies an average individual has shared great-great grandparents with only a small minority.

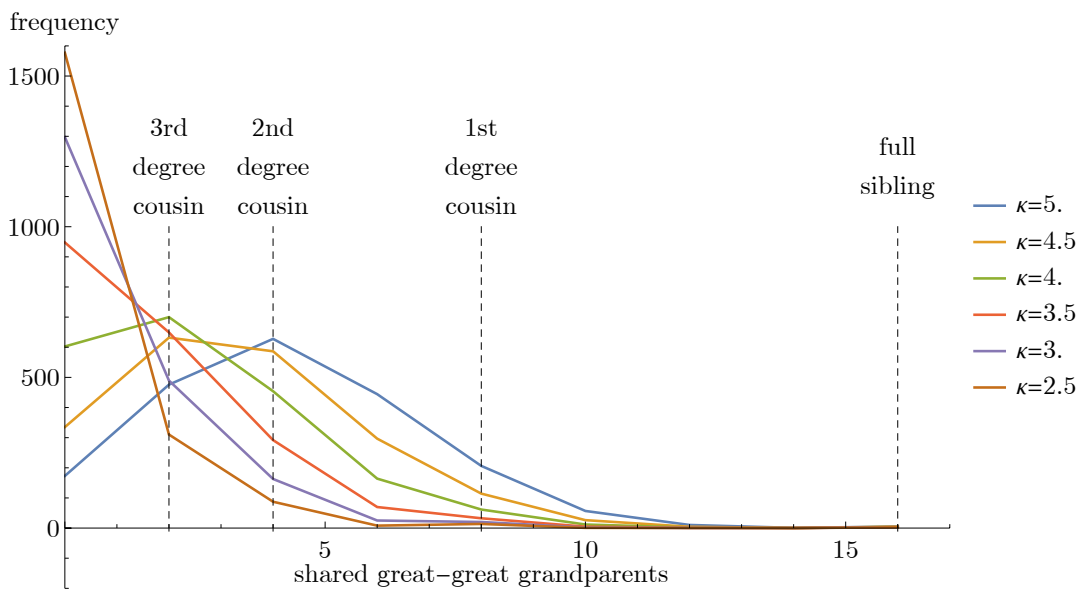

Fig. 5. Relatedness histograms as a function of fertility. (Note that the cousin degree markers are for reference only, as there are other ways to share eight or four great-great grandparents than being 1st and 2nd degree cousins. The group size within which the frequency is measured: 2000. Simulated sample size: 400.)

The number of great-great grandparents that an average group member shares with an average other group member falls from around 4 in high fertility societies to around 0.5 in low fertility societies (Fig. 6). That is, the average relatedness in a high fertility population of 2000 is the same as for 2nd degree cousins, while the average relatedness in a low fertility population much lower than 3rd degree cousins. In other words, the social environment of an average person living in high fertility societies is filled with relatives, while in a low fertility society it is filled with non-relatives.

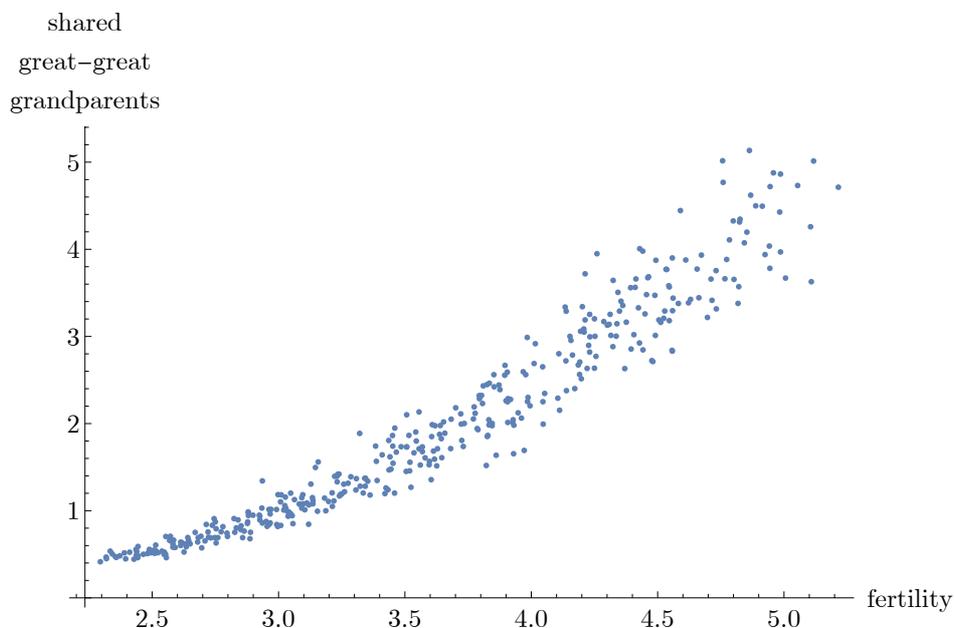

Fig. 6. The number of great-great grandparents shared with an average group member is increasing with the average fertility in the population. (Population size: 2000, n=400.)

**Assessing kinship as a network heuristic.** Thus, linguistic assignment of genetic relatedness covers a large number of individuals in high fertility societies, and only few in low fertility societies. This might explain why kinship terms are used to describe a decreasing section of the population as a society moves from high fertility to low fertility. However, it does not explain the interaction between genetic relatedness and kinship. To capture that, let us focus on how kinship can operate as a network heuristic.

First, I build a kin network made of social network connections only between people who share at least 8 great-great-grandparents. That is, let us start with a social network that consists solely of siblings and first-degree-cousins.

The creation of this kin network allows us to count the number of shared connections between any two people, i.e., the number of other people to whom they are both connected. Fig. 7 shows how the number of shared great-great grandparents affects the number of shared contacts.

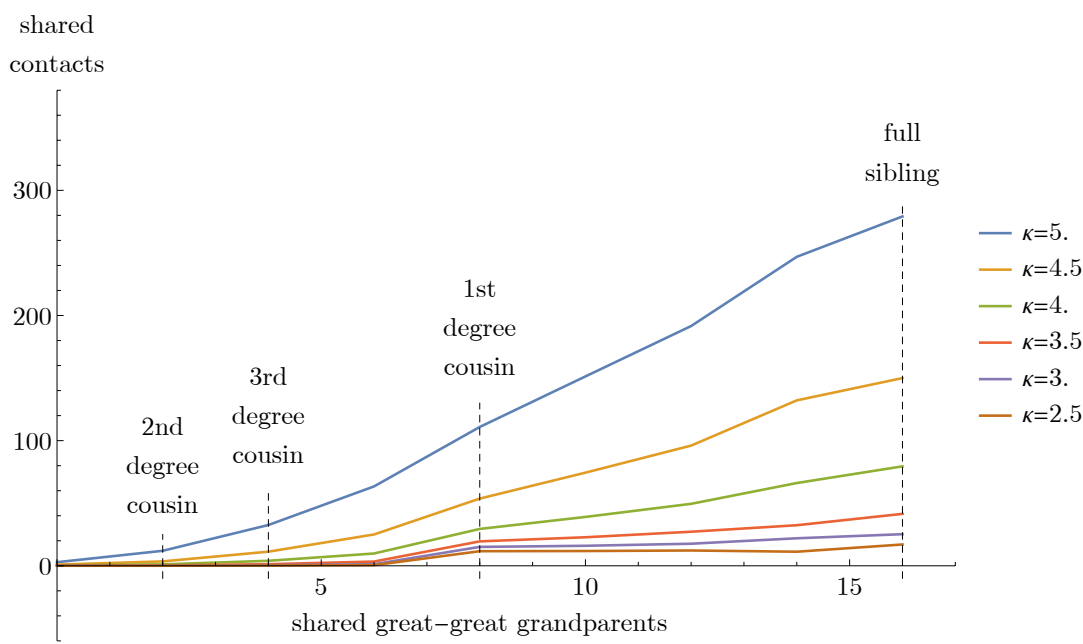

Fig. 7. Kinship degree predicts the shared number of contacts efficiently only in high fertility societies.

As Fig. 7 shows, in the primary, kin-only social network the kin relationship degree, i.e., the number of shared great-great grandparents, predicts the number of shared connections between any two individuals. Thus, when a person (ego) meets another person (alter) with whom there is no social connection, i.e., they are not yet linked in the kin network, knowing their relatedness predicts two pieces of useful information. First, the relatedness is a good proxy variable for the number of channels a gossip might be passed on. The more related they are, the more likely that they have plenty of social contacts, and hence if one cheated on the other, many social contacts of the cheater would hear about it. Second, the relatedness also inform whether it would be useful to establish a bond by predicting how much creating a new link between the ego and the alter would increase the number of closed triangles in the ego's immediate social network, and thus increase her social embeddedness.

Notice that in Fig. 7, the number of shared connections increases with relatedness independent of the fertility level. (That is, all the curves are sloping upwards.) The more related two people are the more likely they have relatives in common. However, while this is true for all the curves, the results also show that fertility affects the usefulness of knowing about the shared relatives. In high fertility societies (top line in Fig. 7), relatedness is an efficient predictor of the number of shared contacts. The slope of the curve is high: moving along the x-axis, the number of shared contacts increases substantially. The opposite is true for the low fertility case (the bottom line in Fig. 7). In this case the slope is almost flat: moving along the x-axis does not change the number of shared contacts much.

When cooperation decisions are made based on reputation, it is the information about a potential cooperation partner's past behaviour that guides the expectation of the future behaviour [51-53]. In a social network, reputation does not need to be acquired and maintained through direct interaction with every other person, instead it can spread via third party information: gossip [39, 54, 55].

If a person A is trying to decide whether a previously unmet person B is likely to have cooperative stance towards her, then A will know that if there are many channels of gossip between her and person B, then person B has less incentive to cheat. This is why the number of closed social network triangles is a useful predictor for whether an interaction partner is likely to take a cooperating or cheating stance [39, 40, 42]. If there are many shared connections then cooperating improves the other individual's social network reputation. Similarly, if there are no, or only few shared social network connections, then the cheating stance's relative benefits are higher.

This is the mechanism in which fertility matters for kinship accounting. If person A lives in a high fertility society, to decide whether she should trust person B, she can rely on kinship: a degree of relatedness is a good predictor of the number of shared connections, and thus whether the interaction stance of the other is cooperative or not. In such a society, an individual would benefit to choose to interact with more closely related others when they have a chance, as these partners are more likely to cooperate. And because it is relatedness predicts networks connections, it makes sense to track it. It makes sense to use kinship system.

Notice that this relationship between the degree of genetic relatedness and the likely cooperation stance is entirely due to relative network positions and has nothing to do with inclusive fitness. In this sense, kinship labelling is a social network-based predictor of cooperation, and not an evolutionary phenomenon.

If person A, however, lives in a low fertility society then relatedness is a poor predictor of the number of shared social connections. In other words, an individual living in a low fertility society would be served badly by using the kinship degree to predict the cooperative stance of another individual.

**Kinship degree versus trait similarity.** When fertility falls, the family size decreases, and there are fewer people to cooperate with. When this happens (or due to the similar effect of urbanisation or migration) people replace relatives with friends [40, 42]. To allow the comparison between kinship and friendship, I altered the model and converted each population into standardised social networks the following way.

First, I assigned each sibling relationship, i.e., each pair that shares 16 great-great grandparents, as an existing social network edge.

Second, I chose a random subset of first-degree cousins, such that the total number of edges to relatives, i.e., siblings plus the randomly chosen cousins, does not exceed 50 contacts for any individual. Thus, if the number of siblings and first degree cousins is less than the limit, then the individual has a social network connection to all of them, while if the number is higher, then only some of the cousins were added. In high fertility populations, this left most individuals at or near the 50 relatives limit, while in low fertility societies, most individuals were far under the limit.

Third, I assigned a random number between 0 and 360 to each individual to serve as a compass-like neutral trait value. (The variable behaves like a circle: 0 degree is the same as 360 degrees). This variable allows choosing friends based on similarity. It is a neutral trait in the sense that it affects only friendship choice, but nothing else in the simulation. It is like a preference for music that allows people to cluster based-on preferring Mozart or Puccini or Bartók, Nicki Minaj or Bad Bunny or Die Antwoord.

Fourth, I took the existing social network populated by relatives, and added new connections based on the distance in the trait values: each individual formed these friendships with those others whose trait value was the closest. In other words, friendship formation was based on similarity, i.e., homophily. This way I added enough new social connections to each individual such that all individuals ended up with 60 social connections in total.

Thus, in these four steps each population yielded a social network such that each individual has exactly the same number of social connections.

**Decision to cooperate or not.** In this altered social network, made up of both relatives and friends, the number of shared social connections is driven by both fertility and trait similarity. Thus, in this model, there are two alternative heuristics that could help guessing connection strength: degree of kinship vs. similarity of trait.

The results, presented in Fig. 8, show that high fertility societies are radically different from the low fertility ones in terms of which of the two variables (kinship or similarity) is likely to determine the shared number of contacts. In high fertility populations, the kinship degree drives the slope, while in low fertility, trait similarity does.

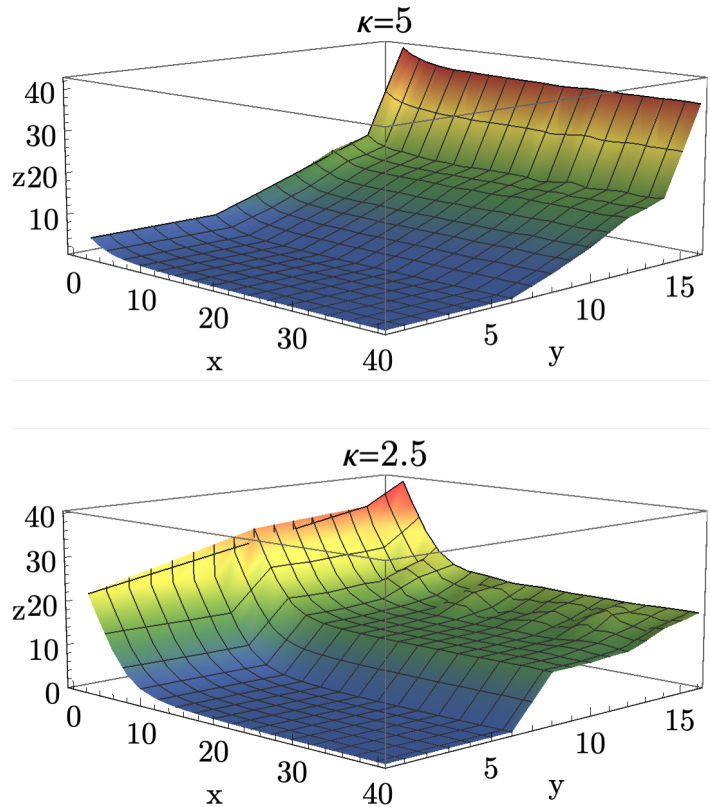

Fig. 8. The interaction between kinship and homophily determining shared number of contacts. x-axis: compass-distance of trait value; y-axis: shared number of great-great grandparents; z-axis: shared number of social contacts. Top panel: average of populations with fertility between 4.5 and 5.5. Bottom panel: average of populations with fertility between 2 and 3.

In a high fertility society (Fig. 8 top panel), the slope of the field increases strongly with relatedness, almost entirely independent of similarity. Which means that knowing the number of shared great-great grandparents helps figuring out the shared social connections with a person (and thus their cooperative stance), while knowing the similarity is not useful.

In a low fertility population (Fig. 8 bottom panel), the slope of the field increases with similarity, almost entirely independent of the shared great-great grandparents (except for the step-change at 8, which is a built-in effect due to the design of the network). Which means that to figure out the cooperativity of a new potential partner, knowing the similarity is important, while the relatedness information is not useful.

To test this effect formally, for each graph I regressed the shared number of social contacts between pairs of individuals on either the pair's shared number of great-great grandparents, or on their compass-distance in trait value. I measured how good predictors relatedness and similarity were using the adjusted-$R^2$ for the (third-degree polynomial) regressions. In other words, I estimated the predictive power of using kinship vs. trait similarity in guessing the number of shared social contacts when meeting a new person.

The results suggest a striking cross-over of the usefulness of tracking kinship degree (Fig. 9). In high fertility societies, tracking kinship is an efficient method for predicting the number of shared social connections, and thus whether an individual should trust an other. In these societies, trait similarity is a useless guide. For low fertility societies the opposite is true: kinship degree is of no use, while trait similarity predicts the shared number of social contacts well.

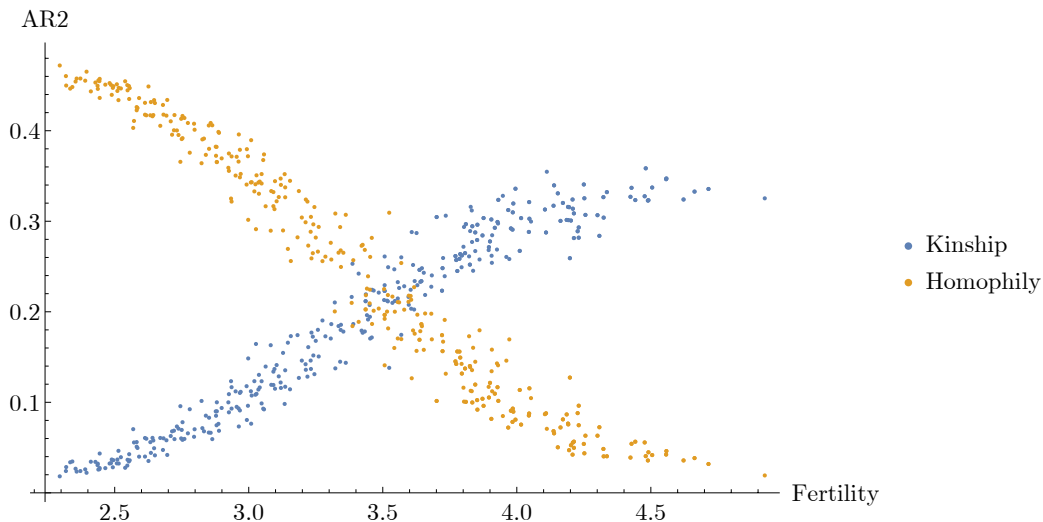

Fig. 9. The predictive power of using kinship and trait similarity in estimating the number of shared social contacts between pairs of individuals. (The y-axis is the predictive power of the model, represented by the adjusted R squared measure. Each fitted curve was third degree polynomial.)

This suggests that kinship accounting is a useful heuristic in high fertility societies, while people living in low fertility societies need to rely on alternative ways to judge the likelihood that a possible partner will take a cooperative stance in their interaction.

## Discussion

In this paper I offer an explanation to the question of why traditional human cultures use kinship labelling, i.e., linguistically assigned degree of genetic relatedness as an organising principle. The results suggest that kinship is not an inclusive fitness phenomenon, but instead it is a social network heuristic.

Notice that the fertility range used in the model is in line with the global societies' change the past 60 years (Fig. 10). Thus, although the result are necessarily stylised as they are produced by simulations, they might inform how the global shift from high fertility to low fertility societies resulted in a shift from predominantly kin-based social networks to predominantly friendship-based ones.

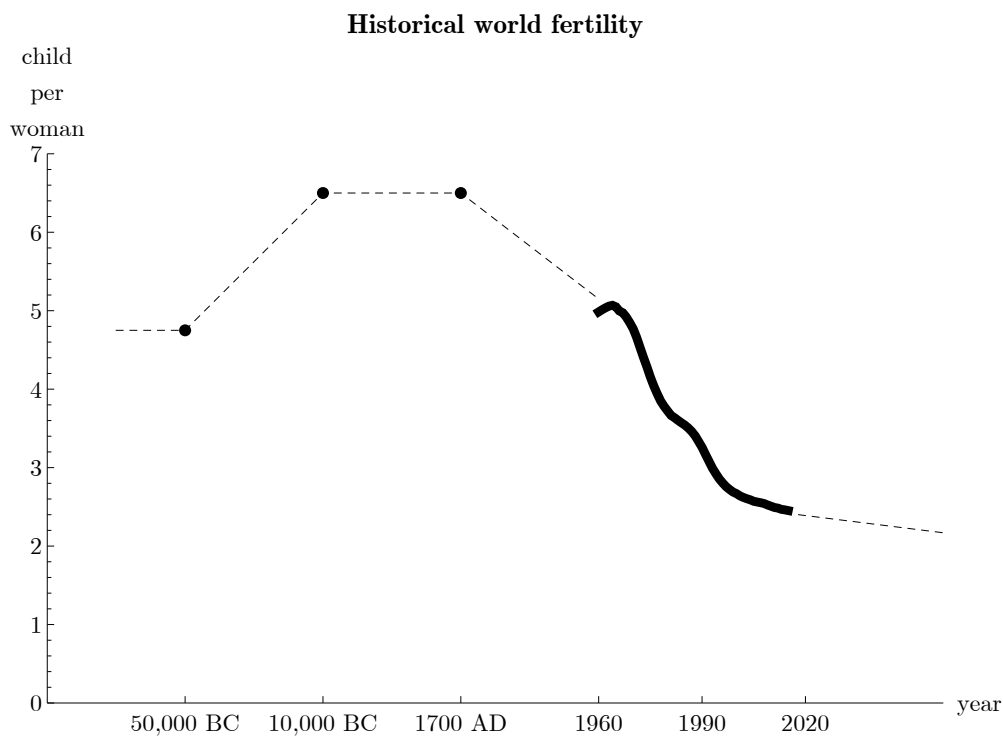

Fig. 10. Global fertility levels. (Dashed line: stylised TFR based on the demographic literature. Continuous line: World Bank database [56].

The model presented in this paper explains five characteristics of kinship that, although have been discussed in the literature of biology, anthropology, and sociology in the past, they have not been previously tackled within a single framework.

**First, the results show why kinship labelling is unique to our species.** The model shows that the degree of genetic relatedness is a useful characteristic upon which cooperate-or-not collaboration stance can be assessed, as long as four basic assumptions are met: (i) the agents have an ability to assign kinship using linguistic terms, (ii) there is a need for collective action in a group that is larger than the immediate family, (iii) a collective practice of tracking kinship is present, and (iv) the effective fertility levels are high enough so that sufficiently large number of relatives are available to populate an average individual's social network.

Although many non-human animals' collective action problems have the characteristics of the (ii)-(iv), despite the occasional presence of "learned vocal labels", e.g., the equivalent of names in bottlenose dolphins [57], no other animal uses categories of kin degree similar to the human linguistic kin assignment. This explains why the phenomenon is unique to humans.

**Second, the results show why kinship is universal among human cultures**. The model demonstrates that if the four above conditions are present, kin degree assignment can be a powerful tool to assess the probability that a new contact will keep to collaboration norms. The most obvious alternative method, i.e., trait similarity-based friendship choice is inferior in efficiency, unless assumption (iv) is violated by not having enough relatives available. This explains why kinship labelling is ubiquitous among human cultures.

**Third, the results show why kinship terminology systems are varied.** The model presented in this paper focuses on the plain vanilla case, in which the social network structure is affected by one or another, i.e., kinship or homophily, as individual-level social network building method. However, even if the above (i)-(iv) characteristics were identically present between two cultures, every factor that might affect the social network would have an impact on the optimal kin tracking system. Dispersal patterns, the structure of the ecological environment, population density, the presence of social hierarchy, and inheritance rules all affect the network structure, and thus all would have impact on the actual manifestation of the network logic of kinship, and thus the optimal kin terminology. This explains the high variation among kinship systems.

**Fourth, the model shows why linguistic kin assignment is imprecise.** Because kin labelling is a network tracking technique, when the factors listed in the previous paragraph impact the social network, the labelling system adjusts, and thus the actual kinship map becomes different to genetic relatedness. This is why kinship systems often track biological relatedness in an imprecise way. The less the close genetic relatedness-based social network is matching the optimal network, the more diversion there should be between the linguistic kin degree and the biological one.

**Fifth, the results show why kinship is replaced by homophily when relatives are scarce.** The model explains why lack of availability of relatives results in a switch from kinship to trait similarity being the guiding social network heuristic.

Notice that while in this model the disappearance of relatives is due to falling fertility, this effect can also come from urbanisation, migration, mortal epidemics or war. Thus, the model provides a theoretical mechanism for the end of kinship as the primary social organisation tool in human societies across the planet, and thus why post-demographic transition societies, urban societies, and populations with frequent migration switched to homophily-based friendship instead.

The switch away from relatives illustrates why the network approach can be useful. If kinship was important for inclusive-fitness-based collective action, then the fall in the number of available relatives would result in reaching out to even more distant relatives. This is clearly not the case in modern societies. Rather than reaching out to third- and fourth-degree cousins, people living in modern, low-fertility, urban societies tend to drop second cousins, sometimes even first cousins, and replace even the few remaining kin with friends.

Thus, kinship is a ubiquitous and unique human trait that, in circumstances that used to be the general case, allows the efficient navigation of social networks of relatives. As such, although the basic dyad is the close family connection recognised in an evolved, inherited way, kinship labelling itself is not an evolutionary phenomenon, but rather a social network tracking heuristic, one that gives way to other tracking methods when the circumstances change.

## Mathematical Methods

This section presents the mathematical model behind the calculations of the paper.

**Building population histories.** Each individual is defined the following way [40]:

$$\{i, g_i, F_{i,-4}, F_{i,-3}, F_{i,-2}, F_{i,-1}\} \tag{1}$$

where $i$ is the index number of the agent, $g_i \in \{0,1\}$ is the gender of the agent, and $F_{i,-4}$ to $F_{i,-1}$ are the index number sets of the great-great grandparents, great grandparents, grandparents, and parents of the agent.

Let $I_s$ denote the set of agents belonging to generation $s$. In each generation, the agents form life-long, monogamous, heterosexual pairs, assuming that their relatedness is more distant than first degree cousins.

$$P_s = \{\{a,b\} \mid g_a \neq g_b \text{ and } a,b \in I_s \text{ and } F_{a,-2} \cap F_{b,-2} = \varnothing\} \tag{2}$$

where $P_s$ is thus the set of pairs formed in generation $s$.

Let $\kappa$ denote the average fertility of the society, and $k$ the actual fertility of any couple, which I assume follows Poisson distribution:

$$k_{P_s} \sim \text{Pois}(\kappa) \tag{3}$$

Then the children born to the couples that constituted generation $s$ form the next generation, $s+1$.

**Standardised population library.** For each fertility rate, $\kappa$, I chose the initial population size such that generation 6 would be just above the target population size of 2000. Then I truncated the final population, such that each simulation results in the same population size, with minimal distortion. (I ran all the simulations in Wolfram Mathematica.) Using this method, I built a library of 400 final populations of six generation population histories, with uniformly distributed variation of fertility between 2.5 and 5.0 total fertility rate, and uniform in the final population group size of 2000 individuals.

**The relationship between relatedness, shared contacts, and fertility.** In each of these populations, I generated a primary kin network in which is an edge exists if the two persons share at least 8 great-great grandparents. Formally, let $a$ denote a binomial adjacency matrix in which

$$a_{i,j} = \begin{cases} 1 \text{ if } |F_{i,-4} \cap F_{j,-4}| \geq 8 \\ 0 \text{ otherwise} \end{cases} \quad (4)$$

where $a_{i,j} \in a$ is 1 if individuals $i$ and $j$ recognise each other as primary kin, and 0 otherwise.

Given the adjacency matrix $a$, it is possible to calculate the number of shared contacts between each pairs of individuals. Formally, let $c$ denote the matrix of the shared contact number, such that

$$c_{i,j} = \sum_{i=1}^{n} \sum_{j=1}^{n} a_{i,k} \cdot a_{j,k} \quad (5)$$

where $c_{i,j} \in c$ is the shared contacts between individuals $i$ and $j$.

Fig. 5 depicts the relationship between the $a$ and $c$ matrices.

Given that each $a$ matrix is the result of a particular population history, and that the population histories varied in the fertility rate, it possible represent the relationship between the fertility rate (as measured from the instantiation of the random process that equation (3) defined) and the elements of the $c$ matrix. Fig. 6 depicts this relationship.

It is possible to create a secondary kinship matrix, which measure the degree of kinship. For this, let $r$ denote the relatedness matrix such that

$$r_{i,j} = |F_{i,-4} \cap F_{j,-4}|$$

where $r_{i,j} \in r$ is the number of shared great-great grandparents between individuals $i$ and $j$.

Fig. 7 depicts the relationship between the $r$ and $c$ matrices dependent on the fertility $\kappa$.

**Kinship degree versus trait similarity.** To compare kinship to friendship, I needed to standardise the individual social networks.

The b1 is all siblings, b2 is a random selection of first-degree cousins such that b1+b2 <= 50, and b3 is the selection of non-relatives such that their trait values are the nearest, and b1+b2+b3=60.

Formally, let $b^{(-1)}$ denote the matrix of siblings such that

$$b_{i,j}^{(-1)} = \begin{cases} 1 \text{ if } F_{i,-1} = F_{j,-1} \\ 0 \text{ otherwise} \end{cases}$$

where $b_{i,j}^{(-1)} \in b^{(-1)}$ is 1 if the individuals $i$ and $j$ are siblings, and 0 otherwise.

Let $b^{(-2)}$ denote the matrix of first-degree cousins such that

$$b_{i,j}^{(-2)} = \begin{cases} 1 \text{ if } |F_{i,-2} \cap F_{j,-2}| = 2 \\ 0 \text{ otherwise} \end{cases}$$

where $b_{i,j}^{(-2)} \in b^{(-2)}$ is 1 if the individuals $i$ and $j$ are first-degree cousins, and 0 otherwise.

To create the possibility of friendship, first a trait value needs to be associated with every individual. For this let, $\phi_i$ denote the trait of individual $i$ the following way:

$$\phi_i \sim U(0°, 360°)$$

where the value is on a compass such that $0° = 360°$.

Let $\delta$ denote the trait similarity matrix, defined as a compass distance:

$$\delta_{i,j} = \begin{cases} |\delta_i - \delta_j| \text{ if } |\delta_i - \delta_j| \leq 180° \\ 360° - |\delta_i - \delta_j| \text{ if } |\delta_i - \delta_j| > 180° \end{cases}$$

where $\delta_{i,j} \in \delta$ is the similarity measure between the trait values of individuals $i$ and $j$. The smaller the $\delta$, the more similar they are.

Using the $b^{(-1)}$, $b^{(-2)}$, and $\delta$ matrices, a social network represented by the adjacency matrix $\tilde{a}$ was built the following way:

First, if $b_{i,j}^{(-1)} = 1$ then $\tilde{a}_{i,j} = 1$, i.e., all sibling relationships are recognised.

Second, select the largest random subgraph $\langle b^{(-2)} \rangle$ such that $\left(\langle b^{(-2)} \rangle + b^{-1}\right).1 \leq \{50\}_n$, and if $\langle b_{i,j}^{(-1)} \rangle = 1$ then $\tilde{a}_{i,j} = 1$ where $n$ is the number of individuals in the society, in this case $n = 2000$. That is, if the number of first-degree cousins plus all siblings is lower than 50, then all first-degree cousins are recognised, otherwise a subset of first-degree cousins are chosen such that the total first-degree cousins plus siblings is 50 for any one individual. In other words, anyone with more than 50 relatives will end up with 50 relatives recognised, not all of them, and anyone with less than 50 relatives will end up recognising all of them.

Third, generate a sorted list, such that $\{...\{i,j\},\{x,y\}...\}$ if $\delta_{i,j} < \delta_{x,y}$, i.e., pairs of individuals with lower compass distance to each other are further up the list. Then, starting from the first element of this list, apply the rule that if $\sum_{y=1}^{n} a_{x,y} < 60$ for both $x = i$, and $x = j$, then $\tilde{a}_{i,j} = 1$, where $i$ and $j$ are the index numbers of the pair in the first element of the list. Then, move on to the next pair, until no more pairs can be found to satisfy the condition. This algorithm fills the social contacts up until 60 total connections. (Note that the average degree of the final graph defined by the $\tilde{a}$ adjacency matrix is a little lower than 60, due to computational reasons.)

This way, individuals in high fertility societies are likely to have 50 relatives (containing all their siblings, and random subset of their first cousins) and 10 friends, while individuals in low fertility societies will end up with a small number of relatives and a larger number of friends, e.g., in a society with fertility 2, and individual has 1 sibling and 4 first-degree cousins, all included in the social network, and an additional 55 friends.

Given the final $\tilde{a}$ matrix, the number of social shared social connections between individuals $i$ and $j$, is

$$\sum_{x=1}^{n} \tilde{a}_{i,x} \tilde{a}_{j,x}$$